\documentclass[letterpaper]{IEEEtran}

\usepackage{cite}                        
\usepackage[pdftex]{graphicx}                      
\usepackage[cmex10]{amsmath}      		  
\usepackage{array}                  
\usepackage{mdwmath}    
\usepackage{mdwtab}                         	
\usepackage{amssymb}
\usepackage{comment}
\usepackage[english]{babel}         	

\usepackage{url}

\begin{document}

\markboth{This article has been accepted for publication. DOI: 10.1109/LAWP.2016.2614338, IEEE Antennas and Wireless Propagation Letters}{}

\title{
	\begin{center}
	On the Spectral Efficiency Limits of an \\
	OAM-based Multiplexing Scheme
	\end{center}
	}

\author{Andrea Cagliero and Rossella Gaffoglio %
\thanks{A. Cagliero and R. Gaffoglio are with the Department of Physics, University of Torino, I-10125 Torino, Italy (e-mail: andrea.cagliero@unito.it; rossella.gaffoglio@unito.it).}
\thanks{{\textbf{1536-1225 (c) 2016 IEEE. Personal use is permitted, but republication$/$redistribution requires IEEE permission. See}}
http:$//$www.ieee.org$/$publications$\_$standards$/$publications$/$rights$/$index.html {\textbf{for more information.}}}}

\maketitle

\selectlanguage{english}

\begin{abstract}

As reported in several recent publications, a spatial multiplexing involving the transmission of orthogonal waves carrying Orbital Angular Momentum (OAM) is unable to provide spectral efficiency improvements with respect to the conventional techniques. In this work we emphasize how the limits of an OAM multiple transmission between antenna arrays can be derived from the Shannon capacity formula, taking as a reference the performance of a multiplexing method based on the higher-order channel modes. Our approach clearly indicates that the two techniques offer the same on-axis performance. Conversely, small misalignments in the arrays positions affect the OAM scheme, highlighting the greater robustness of a traditional multiplexing method in the context of radio communications.

\end{abstract}

\selectlanguage{english}

\IEEEpeerreviewmaketitle

\section{Introduction}

\IEEEPARstart{A}{s} a result of the growing spread of broadband services, the recent years have witnessed a strong increase in the demand for spectral resources within the context of wireless communications. To address this problem, the technological evolution in telecommunications has led to the development of new techniques employing different modulation schemes, polarization and spatial/temporal diversity in order to exploit the electromagnetic spectrum more efficiently. In addition to the above mentioned methods, it has been recently proposed a new approach based on the use of waves carrying Orbital Angular Momentum (OAM), with relevant applications both in optics \cite{Allen1992,Gibson2004,Wang2012} and at the radio frequencies \cite{Thide2007,Tamburini2012,Yan2014}. 

OAM beams are well-known solutions to the Helmholtz equation, characterized by the presence of a phase singularity along the propagation axis, which determines a central intensity null and twisted wavefronts. Mathematically, such phase structure is described by a screw dislocation of the form $e^{i\ell\varphi}$, where $\varphi$ is the azimuthal angle, while the index $\ell\in\mathbb{Z}$, related to the orbital angular momentum content of the beam, indicates the number of twists performed by the phase profile around the central point of zero intensity \cite{PadgettAllen2000}. 

The orthogonality among OAM modes with different $\ell$ index led to consider the possibility of exploiting the peculiar phase distribution of such beams for conveying multiple communication subchannels at the same frequency \cite{Tamburini2012,Yan2014}. Although this idea sounds innovative from a physical point of view, it has been shown that an OAM-based transmission can be considered as a particular case of spatial multiplexing \cite{Edfords2012,Tamagnone2012,Zhao2015}. Within this framework, the over-quadratic power decay in the central region has been pointed out as the real drawback to the use of OAM beams \cite{Tamagnone2013}; however, it should be noted that such behaviour also affects all the higher-order free-space modes which are generally used in spatial multiplexing. Rather, a more relevant issue lies in the strong sensitivity of the OAM orthogonality to misalignments, which seems to restrict this kind of transmissions to the on-axis line-of-sight (LOS) communications only. 

In this letter we provide an in-depth analysis on the limits of OAM radio transmissions by proposing a fair comparison between two multiplexing techniques, one based on OAM modes with index $\ell\neq 0$ (OAM Mux) and the other on the channel higher-order singular modes (CM Mux). In a scenario involving antenna arrays, the Shannon spectral efficiency is computed in the two cases as a function of the propagation distance for both on-axis and off-axis configurations; then, a more realistic example is implemented with the use of some commonly employed modulation schemes and the corresponding $C/N$ reference values under Gaussian channel assumption.

\section{Channel matrix}

Let's consider a communication link made of two facing antenna arrays. For an incident electromagnetic beam generated by $N_T$ transmitting antennas, the circuit voltage induced on a $p$th receiving element can be expressed by \cite{Orfanidis}:
\begin{equation} \label{eq_f1}
V^{\substack{\text{{\tiny{R}}}}}_{p}=\frac{ik\eta}{R}\:\sum^{N_{T}}_{n=1}\frac{e^{-ikR_{np}}}{4\pi R_{np}}\:V^{\substack{\text{{\tiny{T}}}}}_{n}\:\mathbf{h}^{\substack{\text{{\tiny{T}}}}}_{np}\cdot\mathbf{h}^{\substack{\text{{\tiny{R}}}}}_{pn} +w_p,
\end{equation}
where:
\begin{equation} \label{eq_f1bis}
\mathbf{E}_{p}=\frac{ik\eta}{R}\:\sum^{N_{T}}_{n=1}\frac{e^{-ikR_{np}}}{4\pi R_{np}}\:V^{\substack{\text{{\tiny{T}}}}}_{n}\:\mathbf{h}^{\substack{\text{{\tiny{T}}}}}_{np}
\end{equation}
represents the electric field evaluated at the spatial location of the $p$th antenna. In the above expressions, $k=2\pi/\lambda$ is the modulus of the wave vector, being $\lambda$ the wavelength, $\eta$ is the vacuum impedance, $R$ is the resistance of the radiators and $V^{\substack{\text{{\tiny{T}}}}}_{n}=V_0\:\xi^{\substack{\text{{\tiny{T}}}}}_{n}$ is the voltage supply relative to the $n$th transmitting antenna, being $\boldsymbol{\xi}^{\substack{\text{{\tiny{T}}}}}\in\mathbb{C}^{N_T}$ a unity-normalized vector of input coefficients and $V_0$ a voltage constant term associated to the input power $P_{in}$ through the following expression:
\begin{equation} \label{eq_f2}
P_{in}=\frac{V^2_0}{2R}\:\sum^{N_T}_{n=1}\left|\xi^{\substack{\text{{\tiny{T}}}}}_{n}\right|^2=\frac{V^2_0}{2R}.
\end{equation}
Furthermore, $R_{np}$ is the modulus of the vector connecting the $n$th transmitting element to the $p$th receiving one, $\mathbf{h}^{\substack{\text{{\tiny{T}}}}}_{np}$ and $\mathbf{h}^{\substack{\text{{\tiny{R}}}}}_{pn}$ are the effective heights for each reciprocal link, evaluated in the reference frame of the $n$th transmitter and the $p$th receiver, respectively, while $w_p$ is the noise contribution. For simplicity, mutual couplings have been neglected.

\begin{figure}[!t]
\centering
\includegraphics[scale=0.3]{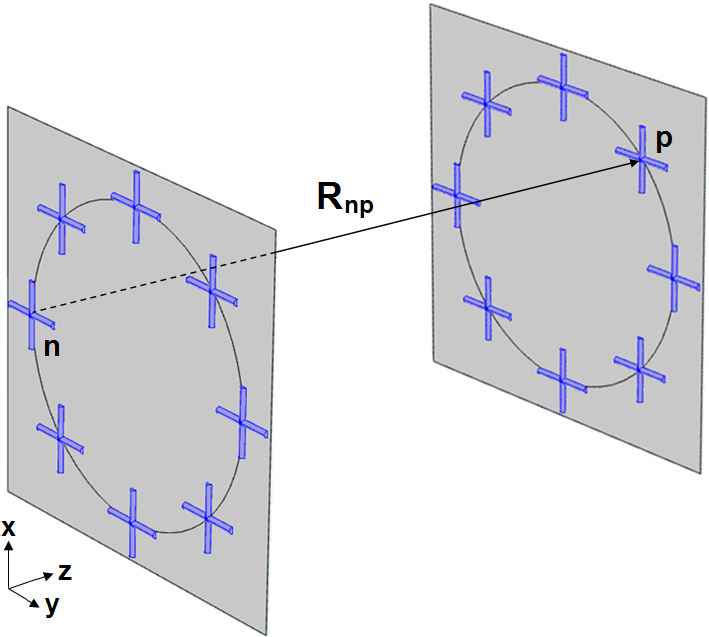}
\caption{Schematic representation of two facing uniform circular arrays each composed by eight crossed half-wave dipoles.}
\label{fig_Gaffo1}
\end{figure} 

\begin{figure}[!t]
\centering
\includegraphics[scale=0.5]{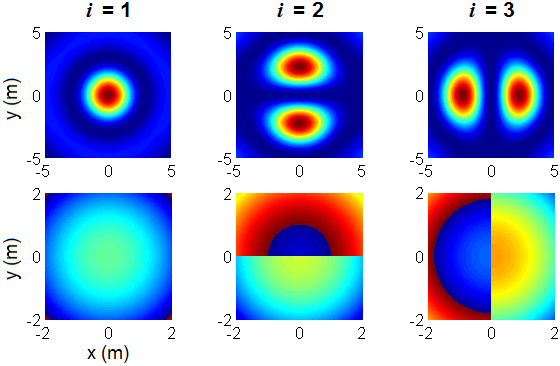}
\caption{Intensity (upper row) and phase (lower row) electric field profiles of the first three channel modes ($i=1,2,3$) displayed in the $xy$ plane at $z=15$ m for the couple of facing UCAs reported in Fig. \ref{fig_Gaffo1} (circular polarization has been considered).}
\label{fig_Gaffo2}
\end{figure}

Assuming that the receiving array has $N_R$ elements, with reference to expression (\ref{eq_f1}) the input/output relationship for the considered MIMO system can be written as follows:
\begin{equation} \label{eq_f3}
\mathbf{V}^{\substack{\text{{\tiny{R}}}}}=\mathbf{H}\mathbf{V}^{\substack{\text{{\tiny{T}}}}} + \mathbf{w},
\end{equation}
being $\mathbf{V}^{\substack{\text{{\tiny{T}}}}}\in\mathbb{C}^{N_T}$ the vector of the $N_T$ inputs $V^{\substack{\text{{\tiny{T}}}}}_{n}$, $\mathbf{V}^{\substack{\text{{\tiny{R}}}}}\in\mathbb{C}^{N_R}$ the vector of the $N_R$ outputs $V^{\substack{\text{{\tiny{R}}}}}_{p}$, $\mathbf{w}\in\mathbb{C}^{N_R}$ the receiver noise vector and $\mathbf{H}\in\mathbb{C}^{N_R\times N_T}$ the channel matrix with elements:
\begin{equation}\label{eq_f4}
H_{pn}=\frac{ik\eta}{R}\:\frac{e^{-ikR_{np}}}{4\pi R_{np}}\:\mathbf{h}^{\substack{\text{{\tiny{T}}}}}_{np}\cdot\mathbf{h}^{\substack{\text{{\tiny{R}}}}}_{pn}. 
\end{equation}
By making use of (\ref{eq_f4}), the total power intercepted by the receiving array in the absence of noise can be defined as:
\begin{equation} \label{eq_f5}
P_{out}=\frac{1}{8R}\:\left|\sum^{N_R}_{p=1}\xi^{\substack{\text{{\tiny{R}}}}}_p\:\sum^{N_T}_{n=1}H_{pn}V^{\substack{\text{{\tiny{T}}}}}_n\right|^2,
\end{equation}
where the coefficients $\xi^{\substack{\text{{\tiny{R}}}}}_p$ belong to a unitary-normalized vector $\boldsymbol{\xi}^{\substack{\text{{\tiny{R}}}}}\in\mathbb{C}^{N_R}$ of ideal beamforming weights introduced at the receiver.

To find the transmit and receive vectors $\boldsymbol{\xi}^{\substack{\text{{\tiny{T}}}}}$ and $\boldsymbol{\xi}^{\substack{\text{{\tiny{R}}}}}$ defining the channel modes of a communication system, it is necessary to resort to the singular value decomposition (SVD) of the channel matrix, given by:
\begin{equation}\label{eq_f6}
\mathbf{H}=\mathbf{U} \boldsymbol{\Sigma} \mathbf{V}^\dagger,
\end{equation}
where $\mathbf{U}\in\mathbb{C}^{N_R\times N_R}$ and $\mathbf{V}\in\mathbb{C}^{N_T\times N_T}$ are the unitary matrices containing the left and right singular vectors of $\mathbf{H}$, respectively, while $\boldsymbol{\Sigma}\in\mathbb{C}^{N_R\times N_T}$ is a rectangular diagonal matrix whose elements are the singular values of $\mathbf{H}$ arranged by decreasing magnitude. The diagonal entries $\Sigma_{ii}$ determine the gains of the independent subchannels identified by the coefficients $\xi^{\substack{\text{{\tiny{T}}}}}_n=V_{ni}$ and $\xi^{\substack{\text{{\tiny{R}}}}}_p=U^\ast_{pi}$.

Fig. \ref{fig_Gaffo1} shows a communication link consisting of a couple of facing uniform circular arrays (UCAs) made of eight crossed half-wave dipoles. For the considered system the feed coefficients $V_{ni}$ are obtained by performing the SVD of the channel matrix (\ref{eq_f4}), where the effective heights can be properly chosen in order to account for the emission of circularly polarized radiation. The corresponding intensity and phase profiles of the electric field (\ref{eq_f1bis}) for the first three circularly polarized channel modes are displayed in Fig. \ref{fig_Gaffo2} for a frequency $f=584$ MHz in the UHF band and a radius $R=1$ m for both the arrays.

\section{OAM modes}

\begin{figure}[!t]
\centering
\includegraphics[scale=0.5]{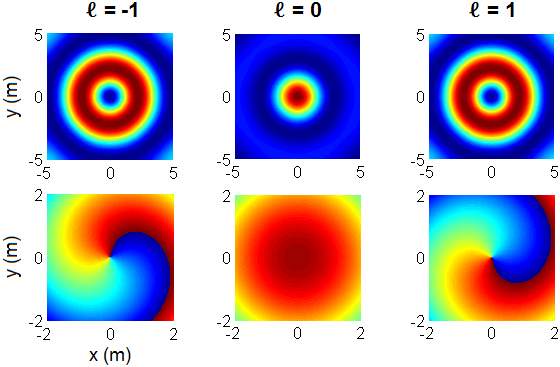}
\caption{Intensity (upper row) and phase (lower row) electric field profiles of the OAM modes with index $\ell=-1,0,1$ displayed in the $xy$ plane at $z=15$ m for an UCA made of eight circularly polarized crossed half-wave dipoles.}
\label{fig_Gaffo3}
\end{figure} 

In the radio domain, OAM beams can be generated by means of an array synthesis process. Starting from given requirements on the array structure, this method allows to find the elements excitations that best reproduce a certain electromagnetic field by solving a linear inverse problem. 

Among all the possible array geometries, the most natural layout for the production of OAM radiation with index $\ell_T$ is that of a $N$-element UCA with $N>2|\ell_T|$ \cite{Thide2007}, which exploits the beams circular symmetry minimizing the number of employed antennas. In this case the input excitations are simply given by:
\begin{equation}\label{eq_f7}
\xi^{\substack{\text{{\tiny{T}}}}}_n=\frac{1}{\sqrt{N}}\:\exp\left[i\ell_T\:2\pi\left(\frac{n-1}{N}\right)\right]
\end{equation}
so that the phase difference between each couple of subsequent elements results $2\pi\ell_T/N$.

When considering a communication channel made of two facing UCAs, the helical phase structure of the incoming OAM beam must be properly compensated in order to avoid a null on-axis power transfer response. To this end, by exploiting the reciprocity theorem and the intrinsic orthogonality of OAM beams, the receiving weight coefficients:
\begin{equation}\label{eq_f8}
\xi^{\substack{\text{{\tiny{R}}}}}_p=\frac{1}{\sqrt{N}}\:\exp\left[-i\ell_R\:2\pi\left(\frac{p-1}{N}\right)\right],
\end{equation}
that must be applied to a $N$-element receiving UCA in order to maximize the on-axis reception, are obtained by imposing $\ell_R=\ell_T$ \cite{Cagliero2016}. Fig. \ref{fig_Gaffo3} shows the intensity and phase electric field profiles of some OAM modes with different $\ell$ index, generated by the transmitting array of Fig. \ref{fig_Gaffo1} at $f=584$ MHz and $R=1$ m. It should be noted that the choice of considering circularly polarized crossed dipoles ensures the circular symmetry and thus the OAM orthogonality to be fully preserved.

\section{Shannon capacity curves}

\begin{figure}[!t]
\centering
\includegraphics[scale=0.6]{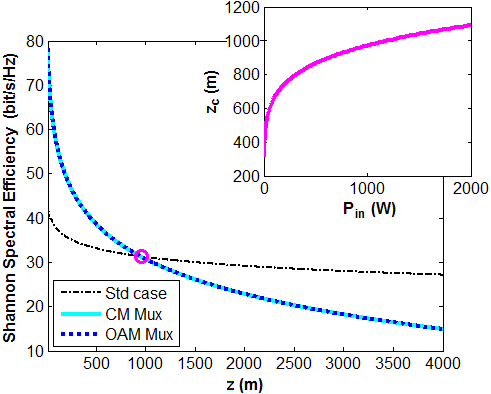}
\caption{Shannon spectral efficiency for two different multiplexing/demultiplexing methods as a function of the link distance between the facing UCAs of Fig. \ref{fig_Gaffo1}, in the absence of misalignments and for a total input power of $1$ kW. The inset shows the critical distance at which the single-channel transmission of the fundamental $\ell=0$ mode (Std case) outperforms the multiplexing methods as a function of the total input power.}
\label{fig_Gaffo4}
\end{figure} 

\begin{figure}[!t]
\centering
\includegraphics[scale=0.6]{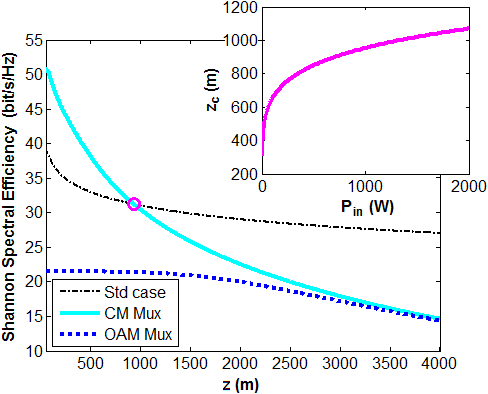}
\caption{Shannon spectral efficiency for two different multiplexing/demultiplexing methods as a function of the link distance, when the total input power is fixed at 1 kW. In this case the receiving array of Fig. \ref{fig_Gaffo1} is rotated by $2$ degrees around the $y$-axis passing through its center (similar results can be obtained for a rotation around the $x$-axis).}
\label{fig_Gaffo5}
\end{figure} 

Making use of the Shannon capacity formula \cite{Guimaraes}, the total spectral efficiency associated to a mode-division multiplexed transmission over a band-limited Gaussian channel can be expressed by:
\begin{equation}\label{eq_f9}
\eta=\sum_{i=1}^Q \log_2 \left[1+P_{out}^{ii}/\left(N_0+\sum\limits_{\substack{j=1,j\ne i}}^Q P_{out}^{ji}\right)\right],
\end{equation}
where $Q$ indicates the number of transmitted modes, $P_{out}^{ii}$ is the properly received power relative to the $i$th subchannel, $N_0$ is the thermal Gaussian noise in the considered bandwidth, while $P_{out}^{ji}$ are noise power contributions representing the crosstalk among the different modes. As a result, each power ratio in (\ref{eq_f9}) can be intended as the signal-to-interference-plus-noise ratio (SINR) of the $i$th subchannel.

In order to provide a comparison in terms of spectral efficiency between two different multiplexing schemes, we considered as a common framework the LOS link depicted in Fig. \ref{fig_Gaffo1} and we set the total input power to a fixed value. Then, we evaluated the total spectral efficiency (\ref{eq_f9}) relative to the simultaneous transmission of the second and the third channel modes (CM Mux), on one hand, and of two OAM beams with index $\ell=\pm 1$ (OAM Mux), on the other, as a function of the link distance. The total input power has been equally divided in both cases between the two considered subchannels and the power contributions $P_{out}$ have been estimated from (\ref{eq_f5}) by means of the {\scshape Matlab}\textsuperscript\textregistered $\!$ software. Given the $i$th subchannel, $P^{ii}_{out}$ in (\ref{eq_f9}) represents the power associated to a mode-matched reception, while the $P^{ji}_{out}$ noise contribution provides the  intercepted power from the transmitted mode $j\neq i$ (or $\ell_T\neq\ell_R$ in the OAM case), which is negligible only in the absence of misalignments for the modes orthogonality. As we can see from Fig. \ref{fig_Gaffo4}, the two multiplexing methods show the same performance in  the on-axis case, i.e., when the arrays are facing each other perfectly. Conversely, as shown in Fig. \ref{fig_Gaffo5}, when a small angular shift in the arrays position is introduced, the CM Mux outperforms the OAM-based scheme, highlighting the greater sensitivity to misalignments of the latter method, due to the symmetry of the beam profiles. The single-channel transmission of the fundamental $\ell=0$ mode (Std case), whose spectral efficiency curve is also reported for comparison, outperforms both the multiplexing methods at a critical distance $z_c$ which depends on the value of the total input power $P_{in}$ (see the insets of Fig. \ref{fig_Gaffo4} and \ref{fig_Gaffo5}). It should be noted that in the off-axis configuration the presented analysis has been limited to cases in which $z_c$ can actually be defined. In both the reported figures the considered arrays have a radius $R=1$ m, the emitted radiation is circularly polarized and the frequency is fixed at $584$ MHz with a channel bandwidth of $7.61$ MHz. Moreover, it is important to emphasize that, even in presence of misalignments, the beamforming coefficients used to determine the $P_{out}$ contributions in the CM Mux case descend from the SVD of the channel matrix relative to the perfectly aligned arrays, ensuring a fair comparison between the two multiplexing methods.

\section{Modulation schemes}

\begin{figure}[!t]
\centering
\includegraphics[scale=0.6]{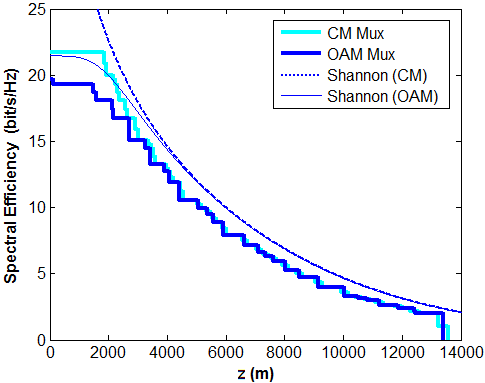}
\caption{Spectral efficiency as a function of the link distance: comparison between the CM Mux and the OAM Mux when an angular shift of $2$ degrees is introduced in the receiving array. The Shannon curves are also reported for reference.}
\label{fig_Gaffo6}
\end{figure}

The Shannon capacity formula (\ref{eq_f9}) represents the theoretical tightest upper limit on the spectral efficiency and has little to do with the values actually reachable by the current modulation schemes. In order to make the above analysis more realistic, we compared algorithmically the SINR of each subchannel at various link distances with the minimum required $C/N$ values for some of the most commonly used modulation schemes, from QPSK 1/2 to 4096-QAM 9/10 \cite{EBU,C2}, under the assumption of a Gaussian channel. This allowed us to provide, for all the different subchannels, an estimation of the maximum distances at which each modulation scheme could be employed, with the corresponding spectral efficiency. Since the transmitted modes are supposed to be independently modulated, the global contribution for the multiple channel is simply given by the sum of all the spectral efficiencies relative to each subchannel. In Fig. \ref{fig_Gaffo6} the obtained spectral efficiency values are hierarchically organized as a function of the corresponding link distances for an off-axis configuration, showing a trend which is found just below the theoretical Shannon curve, as expected. It should be noted that the curve saturation and the sudden drop at small and large distances, respectively, are due to the bounds chosen for the set of modulation schemes. 

Even in this simplified study, the CM Mux proves more robust with respect to misalignments, at least within a certain distance which is related to the total input power. Since the two multiplexing methods show the same performance in the on-axis case, the corresponding plot is not reported here for brevity.

\section{Conclusions}

The spectral efficiency of a communication link between two UCAs composed by crossed dipoles has been studied. The performance of the OAM-based multiplexing has been compared with that of a scheme exploiting the first two higher order singular modes of the considered channel. We have proven how the latter method shows a better behaviour in the presence of misalignments, while the two techniques share the same performance in the on-axis configuration. Our results clearly demonstrate that, apart from the simplicity of the generating methods, no benefit can be expected from the use of the OAM-based multiplexing in the considered LOS scenario.

\section{Acknowledgments}

We would like to express our deep gratitude to Alberto Morello, Bruno Sacco and Assunta De Vita, from the Centre for Research and Technological Innovation, RAI Radiotelevisione Italiana, for proposing the main ideas behind this work and to Prof. Giuseppe Vecchi, from the Polytechnic University of Torino, for all the precious  teachings.


\begin{thebibliography}{10}
\providecommand{\url}[1]{#1}
\csname url@samestyle\endcsname
\providecommand{\newblock}{\relax}
\providecommand{\bibinfo}[2]{#2}
\providecommand{\BIBentrySTDinterwordspacing}{\spaceskip=0pt\relax}
\providecommand{\BIBentryALTinterwordstretchfactor}{4}
\providecommand{\BIBentryALTinterwordspacing}{\spaceskip=\fontdimen2\font plus
\BIBentryALTinterwordstretchfactor\fontdimen3\font minus
  \fontdimen4\font\relax}
\providecommand{\BIBforeignlanguage}[2]{{%
\expandafter\ifx\csname l@#1\endcsname\relax
\typeout{** WARNING: IEEEtran.bst: No hyphenation pattern has been}%
\typeout{** loaded for the language `#1'. Using the pattern for}%
\typeout{** the default language instead.}%
\else
\language=\csname l@#1\endcsname
\fi
#2}}
\providecommand{\BIBdecl}{\relax}
\BIBdecl

\bibitem{Allen1992}
L.~Allen, M.~W. Beijersbergen, R.~J.~C. Spreeuw, and J.~P. Woerdman, ``Orbital
  angular momentum of light and the transformation of {L}aguerre-{G}aussian
  laser modes,'' \emph{Phys. Rev. A}, vol.~45, no.~11, pp. 8185--8189, 1992.

\bibitem{Gibson2004}
G.~Gibson \emph{et~al.}, ``Free-space information transfer using light beams
  carrying orbital angular momentum,'' \emph{Opt. Express}, vol.~12, no.~22,
  pp. 5448--5456, 2004.

\bibitem{Wang2012}
J.~Wang \emph{et~al.}, ``Terabit free-space data transmission employing orbital
  angular momentum multiplexing,'' \emph{Nature Photon.}, vol.~6, no.~7, pp.
  488--496, 2012.

\bibitem{Thide2007}
B.~Thid\'{e} \emph{et~al.}, ``Utilization of photon orbital angular momentum in
  the low-frequency radio domain,'' \emph{Phys. Rev. Lett.}, vol.~99, no.~8,
  2007.

\bibitem{Tamburini2012}
F.~Tamburini \emph{et~al.}, ``Encoding many channels on the same frequency
  through radio vorticity: first experimental test,'' \emph{New J. Phys.},
  vol.~14, 2012, {Art. ID 033001}.

\bibitem{Yan2014}
Y.~Yan \emph{et~al.}, ``High-capacity millimetre-wave communications with
  orbital angular momentum multiplexing,'' \emph{Nature Commun.}, vol.~5, no.
  4876, 2014.

\bibitem{PadgettAllen2000}
M.~Padgett and L.~Allen, ``Light with a twist in its tail,'' \emph{Contemp.
  Phys.}, vol.~41, no.~5, pp. 275--285, 2000.

\bibitem{Edfords2012}
O.~Edfors and A.~J. Johansson, ``Is orbital angular momentum ({OAM}) based
  radio communication an unexploited area?'' \emph{IEEE Trans. Antennas
  Propag.}, vol.~60, no.~2, pp. 1126--1131, 2012.

\bibitem{Tamagnone2012}
M.~Tamagnone, C.~Craeye, and J.~Perruisseau-Carrier, ``Comment on `{E}ncoding
  many channels on the same frequency through radio vorticity: first
  experimental test','' \emph{New J. Phys.}, vol.~14, 2012, {Art. ID 118001}.

\bibitem{Zhao2015}
N.~Zhao \emph{et~al.}, ``Capacity limits of spatially multiplexed free-space
  communication,'' \emph{Nature Photon.}, vol.~9, pp. 822--826, 2015.

\bibitem{Tamagnone2013}
M.~Tamagnone, C.~Craeye, and J.~Perruisseau-Carrier, ``Comment on `{R}eply to
  {C}omment on {``{E}ncoding many channels on the same frequency through radio
  vorticity: first experimental test''}','' \emph{New J. Phys.}, vol.~15, 2013,
  {Art. ID 078001}.

\bibitem{Orfanidis}
\BIBentryALTinterwordspacing
S.~J. Orfanidis, \emph{{Electromagnetic Waves and Antennas}}, 2014. [Online].
  Available: \url{http://www.ece.rutgers.edu/~orfanidi/ewa}
\BIBentrySTDinterwordspacing

\bibitem{Cagliero2016}
A.~Cagliero \emph{et~al.}, ``A new approach to the link budget concept for an
  {OAM} communication link,'' \emph{IEEE Antennas Wireless Propag. Lett.},
  vol.~15, pp. 568--571, 2016.

\bibitem{Guimaraes}
D.~A. Guimar\~{a}es, \emph{{Digital Transmission}}.\hskip 1em plus 0.5em minus
  0.4em\relax Springer, 2009.

\bibitem{EBU}
\emph{Frequency and network planning aspects of DVB-T2}, EBU TECH 3348, version
  4.1.1, 2014.

\bibitem{C2}
\emph{Digital Video Broadcasting (DVB); Implementation Guidelines for a second
  generation digital cable transmission system (DVB-C2)}, ETSI TS 102 991,
  version 1.3.1, 2016.

\end{thebibliography}
\end{document}